\documentclass[aps,floats]{revtex4}
\input epsf
\usepackage{amsmath,amssymb}
\usepackage{graphicx}
\newcommand{\be}{\begin{equation}}
\newcommand{\ee}{\end{equation}}
\newcommand{\Xparm}{X}

\begin{document}

\title{Consequences of a Cosmic Scalar with Kinetic Coupling to Curvature}

\author{Scott F. Daniel\footnote{scott.f.daniel@dartmouth.edu} and Robert R. Caldwell\footnote{robert.r.caldwell@dartmouth.edu}}
\affiliation{ Department of Physics and Astronomy, Dartmouth College, 6127 Wilder 
Laboratory, Hanover, NH 03755 USA}

\date{\today}

\begin{abstract}
The classical gravitational theory of a scalar field with a gradient coupling to the 
Ricci tensor is examined. This is a scalar-vector-tensor gravitational theory, but in the case that the coupling is weak and the scalar evolves like a quintessence field on cosmological time scales, the field equations within the solar system are similar to a vector-tensor theory predicting tightly-constrained preferred-frame effects. In the early universe, it is shown that strong coupling effects can damp the evolution of the scalar field rolling down a potential to help drive an inflationary epoch. In the absence of a potential, the strong coupling effects drive a coasting expansion epoch which ultimately terminates in a sudden singularity. 
\end{abstract}

\maketitle

\section{Introduction}

The boundary between the study of new fields and new gravitation has been blurred in the efforts to
understand dark energy. The observed cosmic acceleration and missing energy problem have lead to
speculation of the existence of a new field which either acts like a present-day inflaton or moderates a departure from Einsteinian gravity.  Consequently, scalar-tensor theories of gravity have enjoyed fresh study as candidates to explain dark energy phenomena. In this context, we propose to examine a novel kinetic coupling between a scalar field and curvature through the Ricci tensor. In particular, we pursue the consequences of gravitational systems described the by Lagrangian
\begin{equation}
\label{lagrangian}
\mathcal{L_{\text{grav}}}=-\frac{R}{16\pi G}+
\frac{1}{2}\nabla_\mu\phi\nabla_\nu\phi\left(g^{\mu\nu}+2 \alpha R^{\mu\nu}\right) - V(\phi),
\end{equation}
wherein gradients of the scalar modulate the strength of gravity. We are motivated in part to find new mechanisms to control the evolution of a cosmic scalar field, for inflationary or quintessence scenarios. This theory introduces a new parameter, $\alpha$, with dimensions of length-squared. Our aim is to study the dynamical behavior and determine the range for $\alpha$ with respect to cosmological and solar system constraints.

A gradient coupling is cause for alarm, as the sign of the coupling term in the Lagrangian is indeterminate. If the coupling term is sufficiently large and negative to overcome the canonical kinetic energy, then negative energies may result. Instabilities may be avoided if $\alpha$ is sufficiently small or if there occurs a separatrix in the equations of motion which prevents the coupling from overcoming the canonical energy term.

The gravitational Lagrangian for scalar-tensor theories, for comparison, is
\begin{equation}
\mathcal{L}_{\text{S-T}} = -\frac{f(R,\phi)}{16 \pi G} + \frac{1}{2}\left(\nabla\phi\right)^2 - V(\phi).
\end{equation}
To make a correspondence to the Brans-Dicke theory, we set $V=0$ and identify the Brans-Dicke scalar as $\Phi_{BD} = \frac{2 \pi G}{\omega_{BD}}\phi^2$, with $f(R,\phi) = \Phi_{BD} R$ \cite{Will:1993ns}. Another class of scalar-tensor theories dispenses with the scalar field $\phi$ altogether and considers a gravitational Lagrangian consisting of a function of the Ricci scalar only, $f(R)$. (See Refs.~\cite{Carroll:2003wy,Amendola:2006we} for a capsule view of this subfield.) Our theory looks like a simple extension of Brans-Dicke to include a vector interaction. Next, a broad class of vector-tensor theories are described by the Lagrangian
\begin{equation}
 \mathcal{L}_{\text{V-T}} = -\frac{1}{16 \pi G}\bigg[ a_1 R + a_2 K_\mu K^\mu R + a_3 K^\mu K^\nu R_{\mu\nu} + \nabla_\mu K^\nu \nabla_\sigma K^\tau (a_4 g^{\mu\sigma} g_{\nu\tau} + 
a_5 g^\mu_\tau g^\sigma_\nu + a_6 g^\nu_\mu g^\sigma_\tau)\bigg]
\label{eqn:vt}
\end{equation}
(See Ref.~\cite{Will:1993ns} for a summary of the properties of scalar- and vector-tensor theories.)
The theory we aim to study (\ref{lagrangian}) is obviously a hybrid of the scalar- and vector-tensor theories, if we identify $K^\mu = \nabla^\mu\phi$ and set $a_2=a_4=a_5=a_6=0$. In fact, our motivation to study the kinetic coupling to curvature derives in part from recent investigations elsewhere of the 
Einstein-aether theory \cite{Jacobson:2000xp} and scalar-vector-tensor theories \cite{Bekenstein:2004ne} of gravitation.  A kinetic coupling to a cosmic scalar helps define a preferred frame, and introduces new scalar and vector degrees of freedom to gravity. Our theory is different from the two above-cited cases, which enforce a constraint $(\nabla\phi)^2=1$, whereas our scalar obeys a generalized Klein-Gordon equation.

In the following, we study the classical theory described by the Lagrangian $\mathcal{L}_{\text{grav}}$ (\ref{lagrangian}). 
In section \ref{sec:eqns} we give the equations of motion and discuss solution methods.
In section \ref{sec:ppn} we make a parametrized post-Newtonian analysis to show that, for a cosmic scalar field with weak kinetic coupling, the gravitational field equations are similar to that of a vector-tensor theory.
In section \ref{sec:cosmicevol} we analyze the scalar field evolution in the case of strong kinetic coupling during the early universe.
Note that we use metric signature $(+---)$ and curvature convention
${R^{\mu}}_{\nu\sigma\tau} = \partial_\sigma \Gamma^\mu_{\nu\tau} - ...,$ 
and $R_{\nu\tau} = {R^{\mu}}_{\nu\mu\tau}$.

\section{Equations of Motion}
\label{sec:eqns}

The gravitational field equations are obtained by varying the action 
\begin{equation}
S = \int d^4x\, \sqrt{-g} \left( \mathcal{L}_{\text{grav}} +  \mathcal{L}_{\text{matter}} \right)
\end{equation}
with respect to the metric $g_{\mu\nu}$, whereby
\begin{eqnarray}
\label{geom}
R^{\mu\nu} - \frac{1}{2} g^{\mu\nu} R&=&8\pi G_N\left(T^{(m)\mu\nu} +  T^{(\phi)\mu\nu} + T^{(\alpha)\mu\nu}\right) \cr
T^{(\phi)\mu\nu} &=& \nabla^\mu\phi\nabla^\nu\phi-g^{\mu\nu}
\left(\frac{1}{2}\nabla^\gamma\phi\nabla_\gamma\phi-V\right) \cr
T^{(\alpha)\mu\nu} &=& 2\alpha\bigg[R^{\nu\gamma}\nabla^\mu\phi\nabla_\gamma\phi + R^{\mu\gamma}\nabla^\nu\phi\nabla_\gamma\phi-\nabla_\gamma\phi\nabla^\gamma\nabla^\mu\nabla^\nu\phi-\nabla^\mu\nabla^\nu\phi\Box\phi\label{tmunu}
\cr
&&
+\frac{1}{2}g^{\mu\nu}\left((\Box\phi)^2+2\nabla^\gamma\phi\nabla_\gamma\Box\phi+\nabla_\gamma\nabla_\beta\phi\nabla^\gamma\nabla^\beta\phi\right)\bigg]. 
\end{eqnarray}
Here $T^{(m)\mu\nu}$ is the stress-energy tensor for all other matter and radiation. Because the matter and radiation stress-energy is conserved independently of the scalar field, and the Bianchi identities still hold, then the scalar field and order-$\alpha$ terms form a conserved system. The divergence $\nabla_{\mu} ( T^{(\phi)\mu\nu} + T^{(\alpha)\mu\nu} )=0$ leads to the scalar field equation of motion
\begin{equation}
\Box\phi+V^\prime
+\alpha\left[\nabla_\mu R \nabla^\mu\phi+2 R^{\mu\nu}\nabla_\mu\nabla_\nu\phi\right]=0.
\label{phieqn}
\end{equation}
The field equations (\ref{geom}) contain third-derivatives of $\phi$, whilst the scalar field equation (\ref{phieqn}) contains third-derivatives of the metric. 

At this point it is useful to identify a weakly-coupled regime in which $\mathcal{O}(\alpha)$ terms are subdominant to other forms of stress-energy, and a strongly-coupled regime in which $\mathcal{O}(\alpha)$ terms dominate. In the weakly-coupled regime, the system of equations can be reduced to second-order, by treating the coupling terms perturbatively. When the coupling terms and the canonical scalar field stress-energy are subdominant to the matter and radiation,
$||T^{(\phi)\mu\nu}||,\, ||T^{(\alpha)\mu\nu}|  \ll ||T^{(m)\mu\nu}||$, the scalar field equation of
motion becomes
\begin{equation}
\Box\phi+V^\prime
\approx -8 \pi G \alpha\left[\nabla_\mu S^{(m)} \nabla^\mu\phi+2  S^{(m)\mu\nu}\nabla_\mu\nabla_\nu\phi\right].
\end{equation}
Here we define
\begin{equation}
S^{(m)\mu\nu} \equiv T^{(m)\mu\nu}  - \frac{1}{2} g^{\mu\nu}  T^{(m)}. 
\end{equation}
Likewise, we can replace the occurrences of the Ricci tensor in $T^{(\alpha)\mu\nu}$ by $8\pi G S^{(m)\mu\nu}$. As a result, the field equations contain at most second-derivatives of the metric. 
These simplifications are well-suited for the case of a cosmic scalar field evolving homogeneously on a cosmic time-scale, as we turn to study gravitation in the solar system.

\section{PPN analysis}
\label{sec:ppn} 

Given the preponderance of experimental data verifying Newtonian gravity in the weak-field limit, 
we analyze the gravitational field equations to first order beyond the Newtonian approximation,
\begin{equation}
g_{00}=1+2U,  \qquad U\equiv-\frac{Gm}{r}.
\end{equation} 
The metric is decomposed as $g_{\mu\nu}=\eta_{\mu\nu}+h_{\mu\nu}$ 
where $\eta_{\mu\nu}$ is the Minkowski metric, $h_{\mu\nu}$ is a perturbation which is at least first order in $(v/c)^2$, and $v$ is the characteristic velocity of bodies in the solar sytem. Perturbations to the scalar field, generated by the stress-energy of the solar system, are written as $\delta\phi$. We will determine the order of $\delta\phi$ by comparing its field equations to $h_{\mu\nu}$.  We adopt the standard conventions of the parameterized post-Newtonian (PPN) analysis of gravitation (see Ref.~\cite{Will:1993ns}), but again note our metric signature. We use the nomenclature  $\mathcal{O}(1)$ to indicate $\sim(v/c)^2$, $\mathcal{O}(1.5)\sim(v/c)^3$ etc.  Hence, we expand the field equations (\ref{geom},\ref{phieqn}) to linear order in $h_{\mu\nu}$ and $\delta\phi$, making assumptions as follows.
First, (C1) we assume that $\phi$ is spatially-homogeneous and evolves on cosmological time-scales which are much longer than the characteristic time-scales of the solar system or laboratory. Consequently, we discard spatial derivatives and second and higher time derivatives of $\phi$. Second, (C2) we assume that the background scalar field is a subdominant source of stress-energy in the solar system, whereby $||T^{(\phi)\mu\nu}||,\, ||T^{(\alpha)\mu\nu}|  \ll ||T^{(m)\mu\nu}||$. Although this is inherently a scalar-vector-tensor theory of gravity, we will soon find that these assumptions render it very similar to a vector-tensor theory.

The perturbed field equations are
\begin{eqnarray}
\frac{\delta R_{\mu\nu}}{8\pi G}&=&\delta S^{(\text{m})}_{\mu\nu} +\partial_\mu\phi \partial_\nu\delta\phi+ \partial_\mu\delta\phi \partial_\nu\phi-h_{\mu\nu}V-\eta_{\mu\nu}V' \delta\phi\nonumber\\
&&+2\alpha\Big[ 8\pi G\left(
\delta S^{(\text{m})}_{\nu 0}\partial_\mu\phi\dot{\phi}
+\delta S^{(\text{m})}_{\mu 0}\partial_\nu\phi \dot{\phi}
-\frac{1}{2}\eta_{\mu\nu}
\delta S^{(\text{m})}_{0 0}\dot{\phi}^2 \right)
\cr&& \qquad
-\dot{\phi}\left(\partial_{\mu\nu}\delta\dot{\phi}-\dot{\phi}\partial_0\delta\Gamma^0_{\mu\nu}\right)
-\frac{1}{2}\eta_{\mu\nu} 
\eta^{\beta\gamma}\dot{\phi}\left(\partial_{\beta\gamma}\delta\dot{\phi}-\dot{\phi}\partial_\beta\delta\Gamma^0_{0\gamma}\right) \Big].  \label{heqn}
\end{eqnarray}
The perturbed scalar field equation of motion becomes
\begin{eqnarray}
\label{eqn:deltaphi}
\Box\delta\phi&=&\eta^{\alpha\beta}\dot{\phi}\delta\Gamma^0_{\alpha\beta}-V^{\prime\prime}\delta\phi
-8\pi G\alpha\bigg[
\dot{\phi} \partial_0\delta S^{\text(m)}
+4\dot{\phi}^2\left(\ddot{\delta\phi}-\dot{\phi}\delta\Gamma^0_{00}\right)\nonumber\\
&&-4\left\{V^{\prime\prime}\delta\phi\dot{\phi}^2+V^{\prime}\left(-h^{00}\dot{\phi}^2+2\dot{\phi}\dot{\delta\phi}\right)+V\left(\Box\delta\phi-\eta^{\mu\nu}\dot{\phi}\delta\Gamma^0_{\mu\nu}\right)\right\}\bigg]\nonumber\\
&&+\delta\mathcal{O}(\alpha^2\phi^3).\label{deltaphi0}
\end{eqnarray}
And we use the gauge conditions
\begin{equation}
\partial_\mu h^\mu_i = \frac{1}{2}\partial_i h, \qquad
\partial_\mu h^\mu_0 = \frac{1}{2}\partial_0h_{00}+\frac{1}{2}\partial_0h.
\end{equation}
To solve for $h_{00}$ and $h_{ij}$ to $\mathcal{O}(1)$, we collect all of the $\mathcal{O}(1)$ terms in equation (\ref{heqn}).  From our constraint (C2) (which implies $\alpha\dot{\phi}^2\lesssim\mathcal{O}(1)$) and equation (\ref{deltaphi0}), we see that $\delta\phi\sim\mathcal{O}(1.5)$, 
and thus does not enter into this first order expansion.  Furthermore, all of the terms proportional to $\alpha$ are also of too high order (being proportional to $\alpha\dot{\phi}^2\sim\mathcal{O}(1)$ times another perturbative term of order at least one).  Therefore, to $\mathcal{O}(1)$, the 0-0 and i-j components of equation (\ref{heqn}) are unchanged from the case of General Relativity, giving  
\begin{equation}
h^{(1)}_{00} = 2U, \qquad
h^{(1)}_{ij} = 2U\delta_{ij}. \label{horder1}
\end{equation}
Therefore, the Newtonian limit is preserved.

The off-diagonal metric components $h_{0i}$ are higher order, $\mathcal{O}(1.5)$, which now includes contributions from the scalar field fluctuations, $\mathcal{O}(\delta\phi)$ \cite{Will:1993ns}. Hence, we must solve equation (\ref{deltaphi0}) to leading order before solving the 0-i components of equation (\ref{heqn}). We rewrite equation (\ref{deltaphi0}) as
\begin{eqnarray}
\Box\delta\phi&=&\eta^{\alpha\beta}\partial_\sigma\phi\delta\Gamma^\sigma_{\alpha\beta}+\mathcal{O}(\ge2)\nonumber\\
\ddot{\delta\phi}-\nabla^2\delta\phi&=&\dot{\phi}\delta\Gamma^0_{00}-\dot{\phi}\delta\Gamma^0_{ii}+\mathcal{O}(\ge2)\nonumber\\
\nabla^2\delta\phi+\mathcal{O}(2.5)&=&-\frac{1}{2}\dot{\phi}\left(\partial_0h_{00}+\partial_0h_{ii}\right)+\dot{\phi}\partial_ih_{0i}+\mathcal{O}(\ge2)\label{deltaphi1}
\end{eqnarray}
where terms proportional to $V$ and its derivatives have been absorbed into $\mathcal{O}(\ge2)$ by assumption (C2), {\it i.e.} if $T^{\phi}$ is subdominant, then $V\sim\mathcal{O}(\ge1)$ and 
\be
V^{\prime\prime}\delta\phi\dot{\phi}^2+V^{\prime}\left(-h^{00}\dot{\phi}^2+2\dot{\phi}\dot{\delta\phi}\right)+V\left(\Box\delta\phi-\eta^{\mu\nu}\dot{\phi}\delta\Gamma^0_{\mu\nu}\right)\sim\mathcal{O}.(\ge2)
\ee
Using the superpotential $\chi$ 
(see equation 4.29 of Ref.~\cite{Will:1993ns}: $\nabla^2\chi = 2 U$) and solutions (\ref{horder1}), we find to leading order
\begin{equation}
\nabla^2\delta\phi=-2\dot{\phi}\partial_0\nabla^2\chi+\dot{\phi}\partial_ih_{0i}.
\end{equation}
Integrating once gives
\begin{equation}
\partial_i\delta\phi=-2\dot{\phi}\partial_{0i}\chi+\dot{\phi}h_{0i}\sim\mathcal{O}(1.5)
\end{equation}
Collecting terms of order 1.5 or less in equation (\ref{heqn}), we have for the 0-i component
\begin{eqnarray}
\frac{\delta R_{0i}}{8\pi G}&=&\delta S^{(\text{m})}_{0i}+\dot{\phi}^2\left(-2\partial_{0i}\chi+h_{0i}\right)+\mathcal{O}(2)\nonumber\\
&=&-\rho v_i+\dot{\phi}^2(-2\partial_{0i}\chi+h_{0i})\label{zeroi}
\end{eqnarray}
By our assumption (C2),
and the observation that $\nabla^{-2}\rho\sim\mathcal{O}(1)$,
we find that the second term on the right hand side of equation (\ref{zeroi}) will result in a contribution of $\mathcal{O}(2.5)$ to $h_{0i}$, 
so that, to post-Newtonian order, $h_{0i}=\frac{7}{2}V_i+\frac{1}{2}W_i$ where $V_i$ and $W_i$ are defined in equation (4.32) of Ref.~\cite{Will:1993ns}.  This is the same result derived in General Relativity.

Finally, we wish to solve for $h_{00}$ again, only this time, expanding (\ref{heqn}) out to $\mathcal{O}(2)$.  Using the stress-energy for a generalized fluid, following \cite{Will:1993ns}, then
\begin{eqnarray}
\delta (S^{(\text{m})}_{\mu\nu})^{(2)}&=&\rho\left(\frac{1}{2}+\frac{\Pi}{2}+v^2+U+\frac{3}{2}\frac{p}{\rho}\right)\nonumber\\
&=&\frac{1}{8\pi G}\left(\nabla^2 U-2\nabla^2\Phi_1-2\nabla^2\Phi_2-\nabla^2\Phi_3-3\nabla^2\Phi_4\right)
\end{eqnarray}
where $\Pi$ is the ratio of proper energy to rest energy, $p$ is the pressure, and the potentials $\Phi_i$ are defined in equation (4.35) of Ref.~\cite{Will:1993ns}. Next, we expand the Ricci curvature to second order,
\begin{eqnarray}
\delta R_{00}(h^2)&=&\delta R_{00}(h)+\frac{1}{2}\left[h^{ij}\partial_{ij}h_{00}-\partial_ih_{00}\partial_ih_{00}+\partial_jh_{00}\left(\partial_ih_{ij}+\frac{1}{2}\partial_jh_{00}\right)\right]\nonumber\\
&=&\delta R_{00}-\nabla^2U^2-4\nabla^2\Phi_2,
\end{eqnarray}
making use of the expressions in the appendix of Ref.~\cite{Barth:1983hb}.
We find that the zero-zero component of (\ref{heqn}) is
\begin{eqnarray}
\frac{1}{16\pi G}\nabla^2h_{00}-\frac{1}{8\pi G}\left(\nabla^2U^2+4\nabla^2\Phi_2\right)
&=&\frac{1}{8\pi G}\left(\nabla^2U-2\nabla^2\Phi_1-2\nabla^2\Phi_2-\nabla^2\Phi_3-3\nabla^2\Phi_4\right)+2\alpha\dot{\phi}^2\nabla^2U\nonumber\\
h_{00}&=&2U(1+16\pi G\alpha\dot{\phi}^2)+2U^2-4\Phi_1+4\Phi_2-2\Phi_3-6\Phi_4\label{h00}
\end{eqnarray}
We absorb the correction to the $2U$ term into a rescaling of Newton's constant
so that, hereafter, the $G$ appearing in the PPN potentials ($\chi,\,V_i,\,W_i,\,\Phi_i$) is the $G$ in equation (\ref{lagrangian}) multiplied by $1+ \Xparm\equiv 1+16\pi G\alpha\dot{\phi}^2$.  
Now
\begin{eqnarray}
h_{ij}&=&\frac{2U\delta_{ij}}{1+ \Xparm}\approx 2U\delta_{ij}(1-\Xparm)\label{ppnij}\\
h_{i0}&=&\frac{7}{2}\frac{V_i}{(1+ \Xparm)}+\frac{1}{2}\frac{W_i}{(1+ \Xparm)}\approx\left(\frac{7}{2}V_i+\frac{1}{2}W_i\right)(1-\Xparm)\label{ppni0}\\
h_{00}&=&2U+\frac{2U^2+4\Phi_2}{(1+ \Xparm)^2}+\frac{1}{1+ \Xparm}\left(-4\Phi_1-2\Phi_3-6\Phi_4\right)\nonumber\\
&&\approx 2U+(2U^2+4\Phi_2)(1-2 \Xparm)+(-4\Phi_1-2\Phi_3-6\Phi_4)(1-\Xparm)\label{ppn00}
\end{eqnarray}
where we have again ignored terms proportional to $V$ on the basis of (C2).  In terms of the standard PPN parameters, we have
\begin{eqnarray}
\gamma&=&1-\Xparm\qquad\beta=1-2 \Xparm\qquad\xi=0\cr
\alpha_1&=&-6 \Xparm\qquad\alpha_2=-\Xparm\qquad\alpha_3=-2 \Xparm\cr
\zeta_1&=&0\qquad\zeta_2=-5 \Xparm\qquad\zeta_3=-\Xparm\qquad\zeta_4=0, 
\label{eqn:ppn}
\end{eqnarray}
as compared to the results for General Relativity
\begin{eqnarray}
\gamma&=&1\qquad\beta=1\qquad\xi=0\cr
\alpha_1&=&0\qquad\alpha_2=0\qquad\alpha_3=0\cr
\zeta_1&=&0\qquad\zeta_2=0\qquad\zeta_3=0\qquad\zeta_4=0. 
\end{eqnarray}
We note that the scalar field terms involving the canonical kinetic energy and the scalar field potential have been assumed or shown to be negligible, contributing at higher orders, so that the theory is essentially General Relativity plus a vector-tensor interaction. However, the PPN parameters derived above (\ref{eqn:ppn}) cannot be easily obtained from the generalized vector-tensor theories analyzed in the literature: in most cases a constraint equation from the conservation of the field equations (equations 5.50, 5.58 in Ref.~\cite{Will:1993ns}), has been employed which is inequivalent to the modified Klein-Gordon equation (\ref{phieqn},\ref{eqn:deltaphi}) which we use. The vector-tensor Lagrangian (\ref{eqn:vt}) does not contain the kinetic term for the scalar field used in our theory (\ref{lagrangian}). Nevertheless, we obtain effects which are characteristic of vector-tensor theories: modification of the degree to which space is curved by mass ($\gamma$), modification of the non-linearity of gravitational superposition ($\beta$), and preferred-frame effects ($\alpha_{1-3}$).

A recent survey of American and Russian radar measurements of planetary motion constrains $\beta-1=(0\pm1)\times 10^{-4}$ \cite{Pitjeva:2005}.
Similarly, radar ranging of the Cassini probe constrains $\gamma-1=(2.1\pm 2.3)\times 10^{-5}$  \cite{Bertotti:2003}.  These constraints are consistent with our assumption (C2).  However, they are not the tightest constraint on our theory.  By predicting non-zero $\alpha_i$, our PPN analysis shows that our gravitational Lagrangian $\mathcal{L}_{grav}$ (\ref{lagrangian}) specifies a preferred reference frame defined by the four-vector $\nabla^\mu\phi$.  Non-zero $\alpha_3$ also threatens non-conservation of momentum \cite{Bell:1996}.  By considering the orbits of millisecond pulsars in our galaxy, Ref.~\cite{Bell:1996} finds the constraint $|\alpha_3|<2.2\times 10^{-20}$ to $90 \%$ confidence.  Consequently, the predicted PPN parameters  in this theory can differ from General Relativity by terms which are at most of order $10^{-20}$, thereby posing a serious challenge to the theory $\mathcal{L}_{grav}$ (\ref{lagrangian}).
   
\section{Cosmological evolution}
\label{sec:cosmicevol}

The equation of motion for the cosmic scalar field $\phi$ in a spatially-flat Robertson-Walker metric is
\begin{eqnarray}
\ddot{\phi}+3\dot{\phi}\frac{\dot{a}}{a}+V^\prime-2\alpha\left(3\ddot{\phi}\frac{\ddot{a}}{a}+3\dot{\phi}\frac{\dddot{a}}{a}+6\dot{\phi}\frac{\ddot{a}}{a}\frac{\dot{a}}{a}\right)=0.
\label{eqn:cosmicphi}
\end{eqnarray}
The energy density and pressure are
\begin{eqnarray}\rho_\phi&\equiv& T^{(\phi)0}_0 +T^{(\alpha)0}_0= \frac{1}{2}\dot{\phi}^2+V+6\alpha\left(\dot{\phi}^2\frac{\dot{a}^2}{a^2}+\dot{\phi}\ddot{\phi}\frac{\dot{a}}{a}-\dot{\phi}^2\frac{\ddot{a}}{a}\right)\cr
p_\phi&\equiv&\frac{1}{3}\left(T^{(\phi)i}_i +T^{(\alpha)i}_i\right)= \frac{1}{2}\dot{\phi}^2-V-2\alpha\left(4\dot{\phi}\ddot{\phi}\frac{\dot{a}}{a}+2\dot{\phi}^2\frac{\ddot{a}}{a}+\dot{\phi}^2\frac{\dot{a}^2}{a^2}+\ddot{\phi}^2+\dot{\phi}\times\dddot{\phi}\right). 
\end{eqnarray}
However, this system is not universally stable.
If we describe the net  stress-energy as a perfect fluid with an equation of state $w$, then small fluctuations obey the equation
\begin{equation}
c_1 \delta\ddot{\phi} + 3 c_2 H \delta\dot{\phi} - \frac{c_3}{a^2}\nabla^2\delta\phi + V''\delta\phi = 0, 
\end{equation}
where the coefficients are  
\begin{equation}
c_1 = 1 + 3 \alpha H^2 (1 + 3 w), \quad
c_2 = 1 - 3 \alpha H^2 w(1+3 w), \quad
c_3 = 1 - 3 \alpha H^2 (1-w).
\end{equation}
In the weak-coupling regime, with $\sqrt{|\alpha|} \ll H^{-1}$, the system is stable against the rapid growth of small perturbations.
In the strong-coupling regime, with $\sqrt{|\alpha|} \gg H^{-1}$, the system is unstable for $w> -1/3$ because $c_1$ and $c_3$ have opposite signs. Conversely, stability during the standard radiation- and matter-dominated epochs requires $\sqrt{|\alpha|} < H^{-1}$ throughout. The most constraining bound occurs at the beginning of the radiation epoch. When $w< -1/3$ as occurs during inflation, $c_1$ and $c_3$ have the same sign which ensures stability, regardless of the magnitude of $\alpha$. 
In fact, we note that the coefficient $c_1$ multiplying the second derivative of either $\phi$ or $\delta\phi$ changes sign across $w=-1/3$ in the strong-coupling limit, suggesting the possible existence of a separatrix, marking a boundary across which the scalar field cannot evolve.

Given the stability criteria, we are motivated to consider the role of the kinetically-coupled scalar field during inflation. We consider a scalar field potential $V = \frac{1}{2}m^2\phi^2$ in an inflationary background, with scale-factor $a \propto \exp{(Ht)}$. In this case, the $\phi$-equation becomes
\be 
\ddot\phi + 3 H \dot\phi + \frac{m^2}{1 - 6 \alpha H^2}\phi=0 
\ee
so that the kinetic-coupling renormalizes the mass. Defining $\tilde\alpha \equiv -\alpha H^2$, then in the limit $\tilde\alpha \gg 1$ (L1) the effective mass is dramatically reduced, $m \to m/\sqrt{6\tilde\alpha}$. Hence, the kinetic drag terms observed in equation (\ref{eqn:cosmicphi}) slow the field evolution in a way identical to a flattening of the potential. Taking the limit $\tilde\alpha \gg m^2/H^2$ (L2) then the two solutions to the scalar field evolution are $\phi \propto \exp{(\lambda t)}$ with $\lambda_1 = -3 H$ and $\lambda_2 = -m^2/18\tilde\alpha H$. Whereas the first solution decays rapidly, the second solution describes the slow-roll of the scalar field down the potential. As well, the limits (L1-2) ensure that the energy density and pressure are dominated by the scalar field potential, so that $p_\phi \approx -\rho_\phi$. Hence, a self-consistent inflationary solution is possible.

In the absence of a potential, setting $V=0$, the system also has interesting solutions. We further neglect any background radiation or matter, and recast the field equations as
\begin{eqnarray}
\frac{H'}{H} &=& \frac{v'}{v} - \frac{1}{2\alpha v^2} + \frac{1}{12 \alpha H^2} \cr\cr
v'' &=&-\frac{1}{v}\left(v'^2 + (4 + \frac{H'}{H})v v' + (3 + 2 \frac{H'}{H})v^2 - \frac{v^2}{4 \alpha H^2} -\frac{1}{2 \alpha}(3 + 2 \frac{H'}{H})\right)
\end{eqnarray}
where $v \equiv \sqrt{8\pi G} \dot\phi$, $x\equiv\ln a$, and ${}'=d/dx$. We focus on the strong-coupling regime with $\alpha < 0$ and start the field with $|\alpha| \gg H^2 \gg v^2$, $v>0$, and $w<-1/3$. 
Following the scalar field dynamics numerically, we find that the field approaches the analytic solution
\be
v \approx \sqrt{\frac{1}{2\alpha} + \frac{C} {a^2}}, \qquad a \ll \sqrt{2\alpha C},
\ee 
where $C$ is determined by the initial conditions. This analytic solution corresponds to a coasting universe with $w=-1/3$ and $a \propto t$. The full system, however, never reaches this ideal solution. As $a \to a_c \equiv \sqrt{2\alpha C}$, the full solution grows singular: $v$ vanishes like $\propto\sqrt{\log{a_c/a}}$ whereas $v'$ and $\frac{\ddot a}{a}$ diverge like $\propto1/\sqrt{\log{a_c/a}}$. This indicates a {\it sudden singularity} \cite{Barrow:2004xh,Cattoen:2005dx} of the curvature. As the critical value of the scale factor is approached, in finite time the energy density remains finite and positive but the negative pressure grows divergent, corresponding to an equation of state $w \to -\infty$. It seems clear that the kinetic coupling leads to a dangerous instability.
 
\section{Discussion}

The consequences of a kinetic coupling between a cosmic scalar field and the Ricci curvature tensor have been studied in this work. In the weak coupling regime, the theory is severely constrained by precision tests of General Relativity. Solar system measurements limit the contribution by the coupling to the energy density and pressure of a cosmic scalar field to be smaller than $\sim 10^{-6}$ of the critical density, which eliminates a possible role in a quintessence scenario. Although this theory is a hybrid of scalar- and vector-tensor theories, we find that the theory suffers from the same problems as vector-tensor theories. Hence, the coupling stress-energy density is restricted to be smaller than $\sim 10^{-20}$ due to limits on preferred-frame effects. 

A possible new mechanism to control the rolling of a cosmic scalar field has been our primary motivation to investigate this theory. Indeed, we have found that in the strong coupling regime, with large and negative $\alpha$, during an inflationary epoch, the coupling terms act like a brake on the scalar field evolution and effectively flatten the slope of the scalar field potential. This behavior may suggest a possible role for such a coupling in the early universe. 

The theory also displays dangerous instabilities. In the absence of a potential, in the strong coupling regime, with large and negative $\alpha$, the coupling leads to a sudden singularity in the curvature.
The strong coupling regime is non-viable in standard radiation- or matter-dominated epochs due to an instability to fluctuation growth. These pathologies suggest a strong kinetic coupling may be unsuitable for a healthy, long-lived universe.

\acknowledgements
This work was supported in part by NSF AST-0349213.


\end{document}